\documentstyle[12pt]{article}

\input mssymb			
\def\integers{\Bbb Z}

\advance\voffset by -2.2cm
\advance\hoffset by -2cm
\newtheorem{theorem}{Theorem}

\def\eq{\begin{equation}}
\def\eqe{\end{equation}}
\def\eqa{\begin{eqnarray}}
\def\eqae{\end{eqnarray}}

\def\half{\frac{1}{2}}
\def\quat{\frac{1}{4}}

\def\cplex{{\mathchoice {\setbox0=\hbox{$\displaystyle\rm C$}\hbox{\hbox
to0pt{\kern0.4\wd0\vrule height0.9\ht0\hss}\box0}}
{\setbox0=\hbox{$\textstyle\rm C$}\hbox{\hbox
to0pt{\kern0.4\wd0\vrule height0.9\ht0\hss}\box0}}
{\setbox0=\hbox{$\scriptstyle\rm C$}\hbox{\hbox
to0pt{\kern0.4\wd0\vrule height0.9\ht0\hss}\box0}}
{\setbox0=\hbox{$\scriptscriptstyle\rm C$}\hbox{\hbox
to0pt{\kern0.4\wd0\vrule height0.9\ht0\hss}\box0}}}}

\def\reals{{\mathchoice {\setbox0=\hbox{$\displaystyle\rm R$}\hbox{\hbox
to0pt{\kern0.4\wd0\vrule height0.9\ht0\hss}\box0}}
{\setbox0=\hbox{$\textstyle\rm R$}\hbox{\hbox
to0pt{\kern0.4\wd0\vrule height0.9\ht0\hss}\box0}}
{\setbox0=\hbox{$\scriptstyle\rm R$}\hbox{\hbox
to0pt{\kern0.4\wd0\vrule height0.9\ht0\hss}\box0}}
{\setbox0=\hbox{$\scriptscriptstyle\rm R$}\hbox{\hbox
to0pt{\kern0.4\wd0\vrule height0.9\ht0\hss}\box0}}}}

\def\del{\partial}

\def\a{\alpha}
\def\b{\beta}

\def\d{\delta}
\def\e{\epsilon}		
\def\g{\gamma}

\def\k{\kappa}			
\def\l{\lambda}
\def\m{\mu}
\def\n{\nu}
\def\w{\omega}
\def\r{\rho}			
\def\t{\tau}
\def\u{\upsilon}
\def\x{\xi}

\def\S{\Sigma}

\def\cl{{\cal L}}
\def\cm{{\cal M}}
\def\cu{{\cal U}}

\def\mnsn{\frac{M^{N}}{\S_{N}}}
\def\cpone{\cplex P_{1}}
\def\cpn{\cplex P_{N-1}}
\def\uy{\underline{y}}
\def\tb{\bar{t}}
\def\zb{\bar{z}}
\def\yb{\bar{y}}
\def\xb{\bar{x}}
\def\ub{\bar{\u}}

\def\jac{\Phi^{\ast}}
\def\jacn{\Phi^{\ast}_{N}}

\def\hcell{\cu_{\half}}

\begin{document}
\baselineskip 16pt plus 1pt minus 1pt
\hfill DAMTP 93-30
\vskip 24pt
\centerline{\bf Thermodynamics of Vortices in the Plane}
\vskip 24pt
\centerline{P.A.Shah}
\vskip 12pt
\centerline{and}
\vskip 12pt
\centerline{N.S.Manton}
\vskip 12pt
\centerline{\it Department of Applied Mathematics and Theoretical Physics}
\centerline{\it University of Cambridge}
\centerline{\it Silver Street, Cambridge CB3 9EW, U.K.}
\vskip 50pt
\centerline{\bf Abstract}
\vskip 24pt
The thermodynamics of vortices in the critically coupled abelian Higgs
model, defined on the plane, are investigated by placing $N$ vortices
in a region of the plane with periodic boundary conditions: a torus.
It is noted that the moduli space for $N$ vortices, which is the same
as that of $N$ indistinguishable points on a torus, fibrates into a
$\cpn$ bundle over the Jacobi manifold of the torus. The volume of the
moduli space is a product of the area of the base of this bundle
and the volume of the fibre. These two values are determined by
considering two 2-surfaces in the bundle corresponding to a rigid
motion of a vortex configuration, and a motion around a fixed centre
of mass. The partition function for the vortices is proportional to
the volume of the moduli space, and the equation of state for the
vortices is $P(A-4\pi N)=NT$ in the thermodynamic limit, where $P$ is
the pressure, $A$ the area of the region of the plane occupied by the
vortices,  and $T$ the temperature. There is no phase transition.
\vfill
\vskip -12pt
\hfill July 1993
\eject

\section{Introduction}

Solitons, as topological defects in field theories and models for
elementary particles and magnetic flux tubes in superconductors, have
been attracting a lot of interest recently. Defects such as strings
and monopoles, if they exist, are recognised to have a significant
effect on the dynamics of the early universe \cite{kolb}. Solitons
in Skyrme's model, a low energy limit of QCD, are a good model for
baryons and their low energy interactions \cite{skyrmions}.
\par
In general, dealing with the interactions of solitons requires a
treatment of the full field theory in a non-perturbative regime.
However, in certain theories at low energies, the motion of a
soliton configuration can be approximated by a continuous sequence
of static configurations \cite{geoapprox}, in much the same way
that a moving cinema picture is built from a sequence of still
pictures. The energy functional of the fields stays ``close'' to
its minimum, which occurs on the configuration space of stable
static solutions of the field equations. This approximation rules
out higher energy phenomena, like soliton-antisoliton pair creation, as
there are no stable static configurations containing a mixture of
solitons and antisolitons. The finite dimensional manifold
parametrising static configurations is
referred to as the {\it moduli space}, and the interaction of the solitons
in the field theory is modelled by geodesic motion in this space, with
a metric determined from the field theory Lagrangian. This
approximation is expected to be good for low soliton impact velocities,
becoming exact in the limit of soliton velocities tending to zero.
\par
The moduli space approach to soliton dynamics is possible at a
critical value of a
coupling constant in the Lagrangian, where the equations describing
the minima of the energy functional reduce to first order differential
equations, the Bogomolny equations \cite{bogoeqns}, and there are
no forces between the solitons. This is why the moduli space has no
potential, and the motion on it is geodesic motion. Comparison of the
geodesic motion with numerical simulations of the scattering of
critically coupled vortices shows a good agreement for velocities
up to around $\frac{1}{3}c$.
\par
For many applications, it would be useful to know the thermodynamic
behaviour of a large number of solitons. In a recent paper
by one of us \cite{smv1}, it was shown how to derive the
thermodynamics of vortices in the abelian Higgs model at critical
coupling, with the vortices moving on a 2-sphere. The calculations
were performed analytically, but within the framework of the moduli
space approximation. This method can be applied to vortices moving on
a general orientable compact Riemann surface $S$. The Lagrangian
density of this model is
\eq
-\quat f_{\m\n}f^{\m\n} + \half D_{\m}\phi(D^{\m}\phi)^{\ast} -
\frac{1}{8}(|\phi|^{2} - 1)^{2} \label{eq:lag}
\eqe
$\phi$ is complex scalar Higgs field and
$ D_{\m}\phi = \del_{\m}\phi -ia_{\m}\phi$ , where $a_{\m}$ is the
$U(1)$ gauge potential. $f_{\m\n} = \del_{\m}a_{\n} - \del_{\n}a_{\m}$
is the Maxwell field tensor. Although the Bogomolny
equations describing the fields of minimal potential energy
are first order, explicit solutions are not available.
However, a static solution with energy $N\pi$ and total magnetic flux
\eqa
\int f_{12} = 2\pi N & , & N \in \integers \label{eq:flux}
\eqae
is uniquely determined (up to a gauge transformation)
by specifying the positions of the precisely $N$ zeros of the Higgs
field, which may be identified with the positions of $N$ vortices
\cite{bradlow}. Classically, these vortex positions  are
an unlabelled set; the configuration is unaffected by interchanging
any of the vortices so the moduli space $\cm_{N}$ is
$(S)^{N}/ \S_{N}$ where $\S_{N}$ is the permutation group of $N$
elements.
\par
There is a naturally defined metric on $\cm_{N}$, derived from the
kinetic part of the Lagrangian. Let
$g_{ij}(Q)$ denote this metric, where $Q = \{ Q^{i} \}$ is a set of
arbitrary coordinates. In the moduli space approximation the
energy of moving vortices is $\half \pi g_{ij}(Q)\dot{Q}^{i}\dot{Q}^{j}$.
By introducing conjugate momenta $P_{i} = \pi g_{ij}(Q)
\dot{Q}^{j} $, the energy may be written in the form
\eq
E(P,Q) = \frac{1}{2\pi} g^{ij}(Q)P_{i}P_{j}
\eqe
The statistical mechanics of $N$ vortices at temperature $T$ can be
treated by using a Gibbs distribution, where the partition function is
\eq
Z = \frac{1}{h^{2N}} \int_{\cm_{N}} (dP)(dQ) e^{\frac{-E(P,Q)}{T}}
\eqe
and $h$ is Planck's constant. By doing the Gaussian momentum
integrals, this reduces to
\eq
Z = ( \int_{\cm_{N}} (dQ) \sqrt{det \, g_{ij}} )
 (\frac{2\pi^{2} T}{h^{2}})^{N}
\eqe
The first factor is simply the volume of the moduli space. It is known
that the metric on the moduli space is K\"ahler  and hence the
volume of the space can be determined from the areas of 2-surfaces
in it by using homology and symmetry arguments.
\par
In this paper, we investigate the statistical mechanics
of vortices on a torus rather than on a
sphere, thereby eliminating undesirable curvature effects.
The torus corresponds to a unit cell in the plane $\reals^{2}$
with periodic boundary conditions. One expects that in the
thermodynamic limit $N \rightarrow \infty$ at fixed number density
$n = \frac{N}{A}$, the behaviour of the particles should not
depend on the global topology of the manifold they are defined on,
although this has not been proven.

\section{The fibre bundle structure of the moduli space}
To model a torus $M= S^{1} \times S^{1}$, we choose a unit
rectangular cell in the plane with periodic coordinates $(x_{1},x_{2})$
with ranges $x_{1} \in [0,1), x_{2} \in [0,\a)$, and give the
spacetime the metric
\eq
ds^{2} = dx_{0}^{2} - L^{2}(dx_{1}^{2} + dx_{2}^{2})
\eqe
where $L$ is a constant physical scaling factor; the area
of the torus is then $A= L^{2}\a$. We introduce a complex coordinate
$z = x_{1} + ix_{2}$ . The Lagrangian (\ref{eq:lag})
leads to the Bogomolny equations for critically coupled vortices
\eqa
(D_{1} +iD_{2})\phi & = & 0 \label{eq:bogo} \\
f_{12} + \half L^{2}(|\phi|^{2} - 1) & = & 0 \nonumber
\eqae
Here, $\phi$ and $a_{\m}$ are not strictly speaking functions on the
torus, but rather a section and connection on a $U(1)$ bundle over it.
The total flux is quantised as in (\ref{eq:flux}), and $N$ is the
first Chern number of the bundle.
\par
The moduli space for $N$ vortices on the torus $M$
is $\cm_{N} = M^{N} / \S_{N}$. This makes it appropriate to define the
configuration in terms of its {\it divisor} $d(\phi)$. A divisor of a function
is a formal unordered sum of the zeros (counted positively by multiplicity) and
poles (counted negatively) of the function. The divisor of the Higgs
field of $N$ vortices at positions $\{p_{1}, \ldots ,p_{N} \}$ is
\eq
d(\phi) = \sum_{i=1}^{N} 1 \cdot p_{i}
\eqe
The magnitude of the Higgs field away from
the vortex positions is not important for our purposes. Let us therefore
consider the complexification of the $U(1)$ gauge group, namely $\cplex^{*}$.
Under transformations of the form $\phi \rightarrow
e^{i\a(z,\zb)}\phi$ , where $\a(z,\zb) \in \cplex$, the absolute
magnitude  $|\phi|$ is no longer gauge invariant but the positions
of the zeros of $\phi$, and hence the vortex positions, still are.
Using the enlarged gauge group, we can locally put the fields
in the holomorphic gauge where $a_{\zb} \equiv \half (a_{1} + ia_{2})
= 0$. Then the first Bogomolny equation,
which can be written as $(\del_{\zb}-ia_{\zb})\phi = 0$, reduces to
\eq
\frac{\del \phi}{\del \zb} = 0
\eqe
so $\phi$ becomes an analytic function of $z$. In this gauge we can
express the Higgs field as a product of analytic functions with
precisely one zero each on the torus. These functions
are the {\it theta functions}, which are actually
defined on the universal covering space of the torus, the complex
plane $\cplex$. The torus can be represented as the quotient of
$\cplex$ by the lattice generated by $1$ and $\t = i\a$. The theta
functions are not singled valued on this quotient, but the positions of
the zeros in the complex plane map to a single point when quotiented
by the lattice.
\par
The lattice generated by $1$ and $\t$ defines a normalised choice of
what is known as the {\it Jacobi mapping}. The Jacobi mapping allows
us to define a centre of mass for the vortex configuration (which is not
unambiguously possible on the original torus). It is this property that
allows a fibration of the moduli space into coordinates related to
(but not the same as) the centre of mass and relative positions of the
vortices. This uses the theory of generalised theta functions,
quasiperiodic functions (or ``{\it complex analytic relatively automorphic
functions}'' - carafs for short) which
 can be defined for any orientable Riemann surface of genus $g>0$. Our
discussion of them will be brief and specific to the torus, but the
interested reader is referred to \cite{gunning} for a fuller and more
general discussion.
\par
The Jacobi mapping of an orientable Riemann surface $S$ of
genus $g$ maps $S$ to a complex $g$-torus, and is defined in terms
of the integrals of normalised abelian differentials (of the first
kind). On the torus, there is just one normalised abelian
differential, namely $dz$, whose integral is $z$, so the Jacobi
mapping is essentially the identity. The fundamental periods of $dz$
are $1$ and $\t$. More precisely, let us choose a
base point $p_{0}$ on $M$, and identify it with the origin of
coordinates. We define the Jacobi manifold of the torus $J(M)$ by
\eqa
J(M) = \frac{\cplex}{\cl} & , &
\cl = \{n_{1} + n_{2}\t : n_{1},n_{2} \in \integers \}
\eqae
and the Jacobi mapping $\jac$ by
\eqa
\jac : M \mapsto J(M) \\
\jac(x_{1},x_{2}) = x_{1} + ix_{2} & , &
x_{1} \in [0,1), x_{2} \in [0,\a) \nonumber
\eqae
Given the Jacobi mapping there is a natural map from an $N$-vortex
configuration to $\cplex$ defined by $t(\phi) =
\sum_{i=1}^{N}\jac(p_{i})$. $\frac{t}{N}$ is then a particular
choice of a centre of mass for the configuration.
\par
The simplest example of a `classical' theta function is
\eq
\theta_{1}(\t,z) = -i \sum_{n=-\infty}^{n=+\infty} (-1)^{n}
e^{(n+\half)^{2}i\pi\t} e^{(2n+1)i\pi z}
\eqe
which has zeros at the origin in the complex
plane and lattice translates of that. Translation of this function
by lattice vectors has the effect of multiplying it by
non-vanishing factors, termed {\it factors of automorphy}:
\eqa
\theta_{1}(\t,z+1) & = &-\theta_{1}(\t,z) \label{eq:facaut} \\
\theta_{1}(\t,z+\t) & = & -\k(\t,z)\theta_{1}(\t,z) \nonumber
\eqae
where $\k(\t,z) = e^{-2\pi iz}e^{-\pi i\t}$ (we shall now drop the $\t$
argument and just write $\theta_{1}(z)$). There are three other
classical theta functions, related to $\theta_{1}(z)$ by
\eq
\theta_{1}(z) =  -\theta_{2}(z+\half)
= -ie^{\frac{i\pi\t}{4}}e^{i\pi z}\theta_{3}(z+\half+\half\t)
= -ie^{\frac{i\pi\t}{4}}e^{i\pi z}\theta_{4}(z+\half\t) \label{eq:thetaid1}
\eqe
Various identities exist between these four functions; the principal
ones that we will use are summarised below:
\eq
\baselineskip 32pt plus 1pt minus 1pt
\begin{tabular}{|c|c|c|c|c|} \hline
& $\theta_{1}(z)$ & $\theta_{2}(z)$ & $\theta_{3}(z)$ & $\theta_{4}(z)$ \\
\hline
1 & -1 & -1 & 1 & 1 \\
\hline
$\t$ & $-\k$ & $\k$ & $\k$ & $-\k$ \\
\hline
\end{tabular}
\baselineskip 16pt plus 1pt minus 1pt
\eqe
\centerline{{\it The factors of automorphy of classical theta functions}}
\eqa      \label{eq:thetaid2}
\theta_{1}(0) \; = \; \theta_{2}(\half) & = &\theta_{3}(\half + \frac{\t}{2})
 \; = \; \theta_{4}(\frac{\t}{2}) \; = \; 0 \\
\theta_{2}^{2}(z)\theta_{4}^{2} & = & \theta_{4}^{2}(z)\theta_{2}^{2} -
 \theta_{1}^{2}(z)\theta_{3}^{2} \nonumber \\
\theta_{3}^{2}(z)\theta_{4}^{2}  & = &  \theta_{4}^{2}(z)\theta_{3}^{2} -
 \theta_{1}^{2}(z)\theta_{2}^{2} \nonumber \\
\theta_{1}(y+z)\theta_{1}(y-z)\theta_{4}^{2}  & = &
\theta_{1}^{2}(y)\theta_{4}^{2}(z) - \theta_{4}^{2}(y)\theta_{1}^{2}(z)
\nonumber \eqae
where $\theta_{i}$ denotes $\theta_{i}(0)$. The
reader is referred to, for example \cite{ww}, for an extensive list of
the properties these functions possess. These functions can be used to
represent divisors on the torus $M$. Although
they are not single-valued on $M$, the factor of automorphy is
non-vanishing, and hence the zeros of $\theta_{i}(z)$ in the complex
plane map to a single point of $M$.
\par
In the holomorphic gauge we can write the Higgs field locally as
\eqa
\phi(z) = \prod_{i=1}^{N} \theta_{1}(z-z_{i}) & , & z_{i}=\jac(p_{i})
\eqae
which is an example of a caraf. Under translation by a lattice element
$\l$, this has factor of automorphy
\eqa
\x_{\phi}(\l,z) = \left\{ \begin{array}{ll}
        (-1)^{N} & \mbox{for $\l=1$} \\
        \prod_{i=1}^{N} -\k(\t,z-z_{i}) & \mbox{ for $\l=\t$}
        \end{array} \right.
\eqae
The characteristic $c(\x)$ of a factor of automorphy $\x$ of a scalar
caraf is equal to the order of the divisor of the function,
i.e. the number of zeros minus the number of poles. Hence
$c(\x_{\phi}) = N$. It is a special case of a general theorem on
factors of automorphy that we can write
\eq
\x_{\phi} (\l,z) = \r_{t}(\l) \x_{p_{0}}^{N}(\l,z)
\eqe
where $t = t(\phi) = \sum_{i=1}^{N} z_{i}$ as defined above.
$\x_{p_{0}}$ denotes the factor of automorphy of $\theta_{1}(z)$, and
\eqa
\r_{t}(1) = 1  &  \r_{t}(\t) = \exp{2\pi i t}
\eqae
By another theorem, the carafs on the torus with factor of automorphy
$\r_{t} \x_{p_{0}}^{N}$ form  at
fixed $t$ a vector space with dimension $N$. Any basis functions
of this space, $f_{i}(t,z)$ ($i=1 \ldots N$)  are quasiperiodic in
$t$ as well as $z$:
\eqa
f_{i}(t,z+\l) & = & \r_{t}(\l) \x_{p_{0}}^{N}(\l,z)f_{i}(t,z)
\label{eq:transz} \\
f_{i}(t+\l,z) & = & \r_{z}(\l) \chi_{ij}(\l,t) f_{j}(t,z) \label{eq:transt}
\eqae
where $\r_{z}(\l)$ is defined identically to $\r_{t}(\l)$. The vector
valued function $\theta(t,z) = (f_{i}(t,z))$ is termed a {\it
generalised theta function} of rank $N$,  with associated
{\it theta factor} $\chi_{ij}(\l,t)$.
\par
We now give a useful basis for these functions on the torus for general
rank $N$, and give  the theta factors for first and second rank.
For rank 1, a basis having the factors of automorphy
(\ref{eq:transz}) and (\ref{eq:transt}) is
\eq
f(t,z) = \theta_{1}(z-t)
\eqe
Using the properties of $\theta_{1}(z)$ listed in (\ref{eq:facaut})
the theta factor can be computed to be
\eqa
\chi(1,t) & = & -1 \\
\chi(\t,t) & = & -e^{-i\pi \t}e^{-2\pi i t} = -\k(\t,t)
\nonumber \eqae
For rank 2, a basis is provided by the independent functions
\eq
f_{i}(t,z) = \{ \theta_{1}^{2}(z-\frac{t}{2}),\theta_{4}^{2}(z-\frac{t}{2})\}
\eqe
The computation of the theta factor utilises identities on the squares
of theta functions, and leads to
\eqa
\chi(1,t) & = & \frac{1}{\theta_{4}^{2}} \left( \begin{array}{cc}
-\theta_{3}^{2} & \theta_{2}^{2} \\  -\theta_{2}^{2} & \theta_{3}^{2}
\end{array} \right) \\
\chi(\t,t) & = & \left( \begin{array}{cc} 0 & -\k(\t,\frac{t}{2}) \\
 -\k(\t,\frac{t}{2}) & 0 \end{array} \right)
\nonumber \eqae
For rank $N$ even, we will find the following basis useful
\eqa \label{eq:base1}
f_{i}(t,z) & = & \{
\theta_{1}^{N},
\theta_{1}^{N-2}\theta_{4}^{2},
\ldots ,
\theta_{4}^{N}, \\  & &
\theta_{1}^{N-3}\theta_{4}\theta_{2}\theta_{3},
\theta_{1}^{N-5}\theta_{4}^{3}\theta_{2}\theta_{3},
\ldots,
\theta_{1}\theta_{4}^{N-3}\theta_{2}\theta_{3}
\}
\nonumber \eqae
where all these theta functions are evaluated at $z-\frac{t}{N}$.
For $N$ odd we shall use
\eqa \label{eq:base2}
f_{i}(t,z) & = & \{
\theta_{1}^{N},
\theta_{1}^{N-2}\theta_{4}^{2},
\ldots ,
\theta_{1}\theta_{4}^{N-1}, \\ & &
\theta_{1}^{N-3}\theta_{4}\theta_{2}\theta_{3},
\theta_{1}^{N-5}\theta_{4}^{3}\theta_{2}\theta_{3},
\ldots,
\theta_{1}^{2}\theta_{4}^{N-4}\theta_{2}\theta_{3},
\theta_{4}^{N-2}\theta_{2}\theta_{3}
\}
\nonumber \eqae
These functions all have the same factors of automorphy, and are
clearly independent, as they have zeros of orders $N, N-2, N-3, \ldots
,2,1,0$ at $z=\frac{t}{N}$, and lattice translates of it. It is interesting
to note the absence of a basis element
with a zero of order $N-1$. If such an independent
function existed, division by $\theta_{1}^{N}(z-\frac{t}{N})$ would lead to an
elliptic function with one simple pole in a unit
cell, but this contradicts a theorem on elliptic functions. Computation
of the theta factors of these bases is in principle possible, but in
practice difficult; we shall anyway not need to use the theta factors in
our computation of the volume of the moduli space.
\par
The relevance of these generalised theta functions and their bases
to the moduli space of $N$ vortices is clear from the
following theorem \cite{gunning} :

\begin{theorem}
For an orientable Riemann surface $S$ of genus $g$, and any integer
$N>2g-1$, the space $\frac{(S)^{N}}{\S_{N}} = \{ d_{\phi}: d_{\phi}=
\sum_{i=1}^{N} 1 \cdot p_{i}$ , $p_{i} \in S \}$ of positive divisors
of order $N$ on the surface can be given the structure
of a fibre bundle with fibre $\cplex P_{N-g}$ and base space $J(S)$,
the Jacobi manifold of $S$. This bundle has a natural projection
\eqa
\jacn & : &\frac{(S)^{N}}{\S_{N}} \mapsto J(S) \\
\jacn(p_{1} \ldots p_{N}) & = &\jac(p_{1}) + \cdots + \jac(p_{N}) \bmod \cl
\nonumber \eqae
where $\cl$ is the lattice of periods of normalised abelian differentials.
\end{theorem}

We can understand this for the torus $M$, with $g=1$, as follows.
$\jacn$ may be identified with $t \bmod \cl$, with $t$ defined as above.
As noted above, this coordinate is related to a choice of centre
of mass of a configuration; its most important property is that it is
invariant under the permutation group acting on the vortices.
For fixed $t$ we have the $N$-dimensional vector space with a basis
given by (\ref{eq:base1}) or (\ref{eq:base2}) above, and a divisor
$d_{\phi}$ determines an element of this vector space via the
intermediary of the function
\eq
f_{\phi}(u,t,z) = \sum_{i=1}^{N} u_{i}f_{i}(t,z)
\eqe
such that $d(f_{\phi}(u,t,z)) = d_{\phi}$ for a choice of $u_{i}$. The $u_{i}$
give homogeneous coordinates in the $\cpn$ fibre, and are only defined
up to a constant
(non-zero) complex multiple, as $cu_{i}$ represents the same divisor as
$u_{i}$.  If $t$ is restricted to lie in one unit cell of the lattice,
the $u_{i}$ are unique modulo such a constant multiple. The $t$
coordinate is unique modulo lattice translations, and translation of
$t$ by a lattice spacing changes the coordinates $u_{i}$ by the appropriate
theta factor (dependent on the choice of caraf basis functions
$f_{i}(t,z)$). More explicitly,  $(t,u_{i}) \rightarrow
(t+\l,u_{j}\chi^{-1}_{ji}(t,\l))$ (where an irrelevant homogeneous factor of
$\r_z^{-1}(\l)$ has been omitted) leads to the same divisor,
as writing $t^{\prime} = t+\l$ and $u_{i}^{\prime} = u_{j}\chi^{-1}_{ji}$
\eqa
f_{\phi}(u^{\prime},t^{\prime},z) & = & \sum u_{i}^{\prime} f_{i}(t+\l,z) \\
 & = & \sum u_{i}^{\prime} \r_{z}(\l) \chi_{ij}(\l,t) f_{j}(t,z)\nonumber \\
 & = & \r_{z}(\l) \sum  u_{j}  f_{j}(t,z) \nonumber \\
 & = & \r_{z}(\l) f_{\phi}(u,t,z) \nonumber
\eqae
Hence $d(f_{\phi}(u^{\prime},t^{\prime},z)) = d(f_{\phi}(u,t,z))$. The theta
factor therefore describes the non-trivial nature of the bundle over $J(M)$.
\par
With the understanding that $t$ will now be restricted to lie in the unit
cell containing the origin in the complex plane, we now seek to
investigate the form of the metric on the bundle, using
the symmetry of the torus.


\section{The metric on the moduli space $\cm_{N}$}
We begin with the most general Hermitian metric on the moduli space,
which has the structure of a fibre bundle as discussed in section 2 :
\eq
ds^{2} = a(t,\u)dt d\tb + b_{\a}(t,\u)dt d\ub_{\a} +
\bar{b}_{\a}(t,\u)d\tb d\u_{\a} + c_{\a\b}(t,\u)d\u_{\a} d\ub_{\b}
\eqe
$\u_{\a}$ are the \underline{in}homogeneous coordinates on the
fibre $\cpn$, $\a,\b = 1 \ldots N-1$ , and $t$ is the base
coordinate which covers $J(M)$ once. $a$ is real and $c_{\a\b}$ is
Hermitian. We note first that from translation symmetry of the
torus, $a,b_{\a},c_{\a\b}$ must be functions of $\u$ only.
Furthermore, we can use the $180^{\circ}$ rotational symmetry of the
torus to eliminate the functions $b_{\a}$. The proper distance
associated with a small displacement $(\d t, \d \u)$ satisfies
\eq
\d s^{2} (t,\u,\d t,\d \u) = \d s^{2} (-t,\u,-\d t,\d \u)
\eqe
and by choosing $\d t$ real so that $\d t = \d\tb$, this implies that
\eq
b_{\a} \d \ub_{\a} + \bar{b}_{\a} \d \u_{\a} = 0
\eqe
Choosing $\d \u_{\a} = i \d_{\a\b}$ implies that $Im(b_{\b}) = 0$, and
choosing  $\d \u_{\a} =  \d_{\a\b}$ implies that $Re(b_{\b}) = 0$,
hence $b_{\b} = 0$ for each $\b$.
\par
It would seem physically reasonable that for a small rigid motion of
the vortex configuration, which would correspond to shifting the $t$
coordinate but staying stationary in the fibre, the distance moved in
the moduli space should not depend on the relative positions of the
vortices. This is confirmed when we apply the result that the moduli
space is K\"ahler. The K\"ahler form is
\eq
\w = \frac{i}{2} (a(\u)dt \wedge d\tb + c_{\a\b}(\u)d\u_{\a} \wedge
d\ub_{\b})		\label{eq:kf}
\eqe
On a K\"ahler manifold $d\w = 0$ , which implies
\eq
\frac{\del a}{\del \u_{\a}} d\u_{\a} \wedge dt \wedge d\tb \, + \frac{\del
a}{\del \ub_{\a}} d\ub_{\a} \wedge dt \wedge d\tb \, + \frac{\del
c_{\a\b}}{\del \u_{\g}} d\u_{\g} \wedge d\u_{\a} \wedge d\ub_{\b} \, +
\frac{\del c_{\a\b}}{\del \ub_{\g}} d\ub_{\g} \wedge d\u_{\a} \wedge
d\ub_{\b} = 0
\eqe
so
\eq
\frac{\del a}{\del \u_{\a}} = \frac{\del a}{\del \ub_{\a}} = 0
\eqe
and therefore $a$ is a constant on the fibre bundle. The resultant
form of the metric is then
\eq
ds^{2} = a dt d\tb + c_{\a\b}(\u)d\u_{\a} d\ub_{\b} \label{eq:pmet}
\eqe
so the volume of the moduli space is a product of the area of the
base $J(M)$ and the volume of the fibre $\cpn$.
\par
Samols \cite{samols} has given a formula for the
metric at a point on the moduli space in terms of the corresponding
solution of the field equations. Writing $f=\log{|\phi|^{2}}$ (in the
original formulation with gauge group $U(1)$), it follows from the
Bogolmolny equations (\ref{eq:bogo}) that $f$ satisfies
\eq
\del^{2}f + L^{2}(1-e^{f}) = 4\pi \sum_{i=1}^{N} \d^{(2)}(z-y_{i})
\eqe
where $y_{i}$ are the vortex positions in $J(M)$. $f$ can be
expanded as a series about each vortex position
\eqa
f(z,\zb) & = & \log{|z-y_{i}|^{2}} + a_{i}(\uy) + \half b_{i}(\uy)(z-y_{i}) +
\half \bar{b}_{i}(\uy)(\zb-\yb_{i}) + \\
& & c_{i}(\uy)(z-y_{i})^{2} + \bar{c}_{i}(\uy)(\zb-\yb_{i})^{2}
+ d_{i}(\uy)(z-y_{i})(\zb-\yb_{i})
+ \ldots
\nonumber \eqae
where $\uy$ denotes the set of all vortex positions. $b_{i}$ measures
how the centres of the contours of $|\phi|^{2}$ drift away from the
vortex position $y_{i}$ as the magnitude of $|\phi|^{2}$ increases
from zero, and this depends on the positions of all the other
vortices.  The metric on the moduli space is then
\eq
ds^{2} = \sum_{i,j=1}^{N} (L^{2}\d_{ij} +
2\frac{\del \bar{b}_{j}}{\del y_{i}})dy_{i}d\yb_{j} \label{eq:smet}
\eqe
and we shall show in the next section how to relate this to (\ref{eq:pmet}).


\section{Collective motion of N coincident vortices}

We now turn to the calculation of the first factor in the volume of
$\cm_{N}$ : the area of the base of the fibre bundle. From
(\ref{eq:pmet}), this is $a\a$, since the range of $t$ is $\{0 \le
Re(t) < 1 \: , \: 0 \le Im(t) < \a \}$. To calculate $a$
we look for a 2-surface of vortex configurations that is
locally orthogonal to the fibre. Pursuing our observation that the
base coordinate defines a choice for the centre of mass of the
vortices, such a surface should correspond to a rigid
motion of the vortices. A simple choice of such a surface
is obtained by placing all the $N$ vortices at one point,
and allowing that point to vary around the torus. We define
\eq
\cm_{c} = \{ d(\phi) : d(\phi) = N \cdot p \, , \, p \in M \} \subset\mnsn
\eqe
We can express $\cm_{c}$ algebraically in terms of the fibre bundle
coordinates as follows. Write the Higgs field in the holomorphic
gauge as
\eq
\phi (z) = \theta_{1}^{N} (z-y)
\eqe
where $y = \jac (p) \in J(M)$. Recall the basis of carafs of rank $N$
given in (\ref{eq:base1}) and (\ref{eq:base2}),
$f_{i} = \{\theta_{1}^{N} (z-\frac{t}{N}) , \ldots \} $,
where now $t = Ny \bmod \cl$. Keeping $y$ restricted to the small
patch of the unit cell delimited by $0 \le Re(y) < \frac{1}{N}\, ,\, 0 \le
Im(y) < \frac{\a}{N}$ , which we shall denote $\cu_{11}$ (we shall use
$\cu_{rs},\, r,s=1 \ldots N$ to mean the small patch $\frac{r-1}{N} <
Re(y) < \frac{r}{N} , \, \frac{(s-1)\a}{N} < Im(y) < \frac{s\a}{N}$
), then
\eq
\phi (z) = f_{1}(Ny,z)
\eqe
Hence the coordinates of $\cm_{c}$ for $y \in \cu_{11}$ are
\eq
\cm_{c} = \{(t,u): (Ny,1,0 \ldots 0) \, ,\, t\in J(M) \, ,\, u\in\cpn \}
\eqe
This surface is locally orthogonal to the fibre since $u$ is constant,
and as $y$ varies
over $\cu_{11}$, $t$ varies over the whole of $J(M)$. To find the
coordinates of the surface for $y \in \cu_{rs}$ one must apply the
quasiperiodicity of the caraf basis in the $t$ coordinate, corresponding
to the non-trivial structure of the fibre bundle. For $y$
varying over the whole of $J(M)$, the surface intersects each fibre
$N^{2}$ times and always orthogonally. For example, for $y \in \cu_{11}$,
equation (\ref{eq:transt}) implies
\eq
u_{i}(y + \frac{1}{N}) = \chi_{ij}(1,Ny)u_{j}(y)
\eqe
where the LHS are the coordinates in $\cu_{21}$, and a homogenous
factor has been omitted. Using $N =2$ as an example, we have
determined the $\chi(\l,t)$ matrices and hence the surface $\cm_{c}$
has coordinates
\[
(t,u) = \left\{ \begin{array}{ll}
	(2y,1,0) & \mbox{$y \in \cu_{11}$} \\
	(2y-1,\theta_{3}^{2},\theta_{2}^{2}) & \mbox{$y \in\cu_{21}$} \\
	(2y-\t,0,1) & \mbox{$y \in\cu_{12}$} \\
	(2y-1-\t,\theta_{2}^{2},\theta_{3}^{2}) &\mbox{$y\in\cu_{22}$} \\
	\end{array}
\right. \]
Returning to $N$ vortices, the algebraic expression for the
coordinates of $\cm_{c}$ can be
inserted in the metric on the fibre bundle, restricted to $\cm_{c}$. For
$y \in \cu_{11}$, $dt = N dy$ and $d\u = \frac{\del \u}{\del y} dy =
0$ so recalling equation (\ref{eq:pmet})
\eq
ds^{2} \mid_{\cm_{c}} = a N^{2} dy d\yb	\label{eq:dsonmc}
\eqe
\par
We can now find $a$ by determining the metric on $\cm_{c}$ from Samols'
formula (\ref{eq:smet}) which for $N$ coincident vortices reduces to
\eq
ds^{2} \mid_{\cm_{c}} = N(L^{2} + 2\frac{\del b}{\del \yb}) dy d\yb
\label{eq:metforn}
\eqe
where $\half b(y)$ is the linear coefficient in the expansion of
$f=\log{|\phi|^{2}}$ around the vortex position $y$.
The reflection symmetry $z-y \rightarrow -(z-y)$ can be
used in this local expansion, and implies $b(y) = 0$ ; the
contours of the Higgs field are locally centred on the $N$-vortex
position. Hence (\ref{eq:metforn}) simplifies to
\eq
ds^{2} \mid_{\cm_{c}} = NL^{2} dy d\yb
\eqe
which is the same result as would be obtained for $N$ coincident
vortices in a plane, using the more powerful circular symmetry.
Comparison with (\ref{eq:dsonmc}) yields
\eq
a=\frac{L^{2}}{N}
\eqe
Hence
\eq
Area(J(M)) = \frac{L^{2}\a}{N} = \frac{A}{N}	\label{eq:areabase}
\eqe


\section{Simultaneous motion of two diametrically opposite vortices}
In the previous section, we used a rigid collective motion of the
vortex configuration to determine the area of the
base space of the fibre bundle. We now turn to determining the volume
of the fibre, $\cpn$, by looking for a motion that lies entirely in
the fibre, keeping the centre of mass fixed. The simplest example
seems to be the motion where $N-2$ of the vortices are
 fixed at the point $p_{0}$, and the other two are at positions $p_{1}$
and $p_{2}$ symmetrically placed on opposite sides of $p_{0}$ and
moving over some suitable region of the torus. This motion of vortices
corresponds to a motion in a submanifold $\cm_{d}$ of the moduli space,
\eq
\cm_{d} = \{ d(\phi) : d(\phi)=(N-2)\cdot p_{0} + 1\cdot p_{1} +
1\cdot p_{2}, \; p_{1},p_{2} \in M\}
\eqe
where $\jac(p_{1}) = x$, say, and $\jac(p_{2}) = -x \bmod \cl$ , so
the coordinate $t(\phi) = 0$. $\cm_{d}$ is covered once if $x$ ranges
over the half cell $\hcell = \{x: 0\le Re(x) < \half,
0\le Im(x) < \a\}$. As before, we seek to find an expression
for the fibre coordinates on $\cm_{d}$ . Again in the holomorphic gauge
\eq
\phi (z) = \theta_{1}(z-x) \theta_{1}(z+x) \theta_{1}^{N-2}(z)
\eqe
We can use the addition identities of theta
functions (\ref{eq:thetaid2}) to write this as
\eq
\phi (z) = \frac{1}{\theta_{4}^{2}}(\theta_{1}^{N}(z)\theta_{4}^{2}(x)
- \theta_{4}^{2}(z)\theta_{1}^{N-2}(z)\theta_{1}^{2}(x))
\eqe
Recall our basis, with $t=0$ here,
\eq
f_{i}(0,z) = \{\theta_{1}^{N}(z),
\theta_{1}^{N-2}(z)\theta_{4}^{2}(z), \ldots\}
\eqe
Hence, omitting the irrelevant $\theta_{4}^{-2}$ factor, the
base and homogenous fibre coordinates of $\cm_{d}$ are
\eq
\cm_{d} = \{(t,u_{i}) : (0,\theta_{4}^{2}(x),-\theta_{1}^{2}(x),0
\ldots 0) \}
\eqe
Therefore $\cm_{d}$ lies in the $\cpone$ line $u_{3} = u_{4} = \ldots
= u_{N} = 0$. The inhomogenous fibre coordinates are
\eqa
\u_{\a} = \left\{ \begin{array}{ll}
        -\frac{\theta_{1}^{2}(x)}{\theta_{4}^{2}(x)} & \mbox{$\a =1$} \\
        0 & \mbox{otherwise}
        \end{array} \right.
\eqae
which are even elliptic functions of $x$, as expected from the symmetry
of the motion. Now for $x$ varying over a unit cell, the
number of solutions of $\u_{1} = c \in \cplex$ is independent of $c$, and
is equal to the number of zeros or poles of $\u_{1}$, counted by
multiplicity. In this case, there are two solutions, but the symmetry
$\u_{1}(x) = \u_{1}(-x)$ implies that there is exactly one solution in
the half cell $\hcell$. Hence, for $x$ varying over $\hcell$,
$\cm_{d}$ covers a $\cpone$ line of $\cpn$ exactly once.
\par
We now need to apply this result to the metric on the fibre bundle,
whose restriction to $\cm_{d}$ is given by
\eq
ds^{2} |_{\cm_{d}} = 2(L^{2} + \frac{\del b_{x}}{\del \xb} - \frac{\del
b_{-x}}{\del\xb})dx d\xb
\eqe
(remembering that as we vary $x$ we are varying the positions of
two vortices simultaneously). $\half b_{x}$ is the linear coefficient
of the expansion of $f = \log{|\phi^{2}|}$ around the vortex at $x$
and $\half b_{-x}$ the linear coefficient around the vortex at $-x$.
Set $b_{x} = b$, and by $180^{\circ}$ rotational symmetry
$b_{-x} = -b$. However, it is now
not possible to solve for $b$ using the symmetry of
the torus. Fortunately, we do not need to know $b$ everywhere; we can
instead consider the integral of $\frac{\del b(x)}{\del \xb}$ over
$x \in \hcell$, and use Stokes' theorem. The area of $\cm_{d}$ is
\eqa
I & = & \int_{\hcell} (2L^{2} +  2\frac{\del b_{x}}{\del \xb} - 2\frac{\del
b_{-x}}{\del\xb})(\frac{1}{2i} d\xb \wedge dx)
\\
  & = & A - 2i \int_{\hcell} \del_{\xb} (b \: dx)
\nonumber \\
  & = & A + 2i \int_{\del\hcell} b \: dx
\nonumber
\eqae
where the orientation of the contour $\del\hcell$ is shown in figure 1.
$b$ has poles at the points where the vortices at $x$ and $-x$
coincide, and these are marked.
Consider the integral on the two line segments
$l_{1} = \{x: x = c\t , c \in (\e,\half-\e) \cup (\half+\e,1-\e) \}$.
For $x \in l_{1}$, $|\phi(z)|= |\phi(-z)|= |\phi(\zb)|= |\phi(-\zb)|$,
and
\eq
\int_{l_{1}} b \: dx = 0
\eqe
as the parts from $c<\half$ and $c>\half$ cancel. In a similar way, the
integral on $l_{3}$ (see figure) is zero, and $l_{2}$ and $l_{4}$ are
lattice translations of each other with opposite direction, and so
their contributions to $I$ cancel. The only contribution to $I$
comes from the poles indicated. Consider the pole at $x=0$.
To investigate it, we expand $f$ about $z=0$, assuming $x$ is near $0$.
\eqa
f & = & (N-2) \log{|z|^{2}} + \log{|z+x|^{2}} + \log{|z-x|^{2}} + A(x)
+ \\
  &   & \half B(x)z + \half \bar{B}(x)\zb + C(x)z^{2} +
\bar{C}(x)\zb^{2} + D(x)z\zb + \ldots \nonumber
\eqae
where $A,C,D$ are all regular around $x=0$. The symmetry $z \rightarrow
-z$ implies that $B(x)$, and all other coefficients of terms odd in
$z$, vanish.
Likewise, the symmetry $x \rightarrow -x$ implies $A(x),C(x),D(x)$
are all even functions of $x$. To find the singularity in $b$,
compare this with an expansion about $z=x$
\eqa
f & = & \log{|z-x|^{2}} + a(x) + \half b(x)\,(z-x) +
\half \bar{b}(x)\,(\zb - \xb) + \\
     &   & c(x)\,(z-x)^{2} + \bar{c}(x)\,(\zb -\xb)^{2} +
d(x)\,(z-x)(\zb-\xb) + \ldots
\nonumber \eqae
Using $\log{|z+x|^{2}} = \log{|2x|^2} + \frac{1}{2x}(z-x) +
\frac{1}{2\xb}(\zb-\xb) - \frac{1}{8x^{2}}(z-x)^{2} -
\frac{1}{8\xb^{2}}(\zb-\xb)^{2} + \ldots$ (the expansion is valid for
$|z-x|<|2x|$, which includes $z=0$), and similarily for $\log{|z|^2}$
we obtain
\eqa
a(x) & = & A(x) + \log{|2x|^{2}} + (N-2)\log{|x|^{2}} + C(x)x^{2} +
\bar{C}(x)\xb^2 + D(x)x\xb + O(x^{4}) \\
\half b(x) & = & \frac{1}{2x} + \frac{N-2}{x}+ C(x)2x + D(x)\xb +O(x^{3})
\nonumber \\
c(x) & = & -\frac{1}{8x^{2}} -\frac{N-2}{2x^{2}}+ C(x) + O(x^{2})
\nonumber \\
d(x) & = & D(x) + O(x^{2}) \nonumber
\eqae
Hence, $a(x),b(x),c(x)$ have logarithmic, first order and second order
poles at $x=0$, with the singularities completely determined by the
log terms in the expansion around $z=0$. The integral of $2ib$ on the
contour near $x=0$, $l_{\e} = \{x: x=\e e^{i\theta}, \theta \in
(0,\frac{\pi}{2}) \}$ just has a contribution from the pole (the
linear and all higher order terms are odd in $x$ and vanish under
integration):
\eqa
2i\int_{l_{\e}} b \: dx & = & 2i\int_{0}^{\frac{\pi}{2}} (2N-3)i d\theta \\
                      & = & -\pi(2N-3)
\nonumber \eqae
A calculation for the pole of $b$ at
$x=\frac{\t}{2}$, where the two vortices again coincide (but this
time the expansions are unaffected by the other vortices at $z=0$),
yields a contribution to the integral of $-2\pi$ (this time the
integral is around a half circle), and likewise for the pole at $x= \half
+\frac{\t}{2}$. The poles at $x=\half, \half+\t$ contribute $-\pi$
as the integral there is around a quarter circle. Finally, the pole at
$x=\t$ makes the same contribution as $x=0$. The residues of all the
poles of $b$ on $\del \hcell$ are shown in figure 1. Their contributions to
the integral $I$ sum to give
\eqa
I & = & A - 2\pi(2N-3 + 2 + 1) \\
  & = & A - 4\pi N
\nonumber \eqae
The result is always positive, because of the observation in
\cite{bradlow} that the Bogomolny equations for vortices on a
compact Riemann surface only have non-trivial solutions if the
physical area $A$ of the surface is larger than $4\pi N$.
\par
The value of $I$ obtained for $x \in \hcell$ represents the area of a
complex line, $\cpone$, in the fibre $\cpn$ (had we integrated instead
over the whole unit cell $\cu$, we would have obtained double
the result from the sum of pole residues, but would then divide by
two as this surface would have covered $\cpone$ twice).
Now in a complex manifold with a K\"ahler metric, the area of any
complex curve is the integral of the K\"ahler form over it. So
the integral of the K\"ahler form (\ref{eq:kf}) over a complex line
in the fibre is
\eq
     [\w] = \int_{CP_{1}} \w = A - 4 \pi N
\eqe
$\hat{\w} = \w/[\w]$ has integral of unity over
a complex line, and so is a basis element for $H_{2}(\cpn,\integers)$,
the second cohomology group of the fibre over the integers. Then, from
the topology of complex projective space (see e.g. \cite{griffiths}),
$\hat{\w}^{N-1}$ is a basis element for $H_{2N-2}(\cpn,\integers)$.
Now the volume of a K\"ahler manifold of dimension $2N-2$ is the
integral of $\w^{N-1}/(N-1)!$ over it. Therefore the fibre volume is
\eq
     Vol(\cpn) = \frac{(A - 4 \pi N)^{N-1}}{(N-1)!} \label{eq:volcpn}
\eqe


\section{The partition function of N vortices}
We can now easily combine our results for the area of the base space
of the fibre bundle (\ref{eq:areabase}) and the volume of the $\cpn$
fibre (\ref{eq:volcpn}); their product gives the volume of the moduli space
\eq
     Vol(\cm_{N}) = \frac{A}{N!}(A - 4 \pi N)^{N-1} =
\frac{A^{N}}{N!}(1-4\pi n )^{N-1}
\eqe
where $n=\frac{N}{A}$ is the vortex number density. The partition
function is then
\eq
     Z_{torus} = \frac{A^{N}}{N!}(1 - 4 \pi n)^{N-1} (\frac{2 \pi^{2}T}
{h^{2}})^{N}
\eqe
This should be compared with the result obtained in \cite{smv1}
\eq
     Z_{sphere} = \frac{A^{N}}{N!}(1 - 4 \pi n)^{N} (\frac{2 \pi^{2}T}
{h^{2}})^{N}
\eqe
The additional single factor of $(1-4\pi n)$ is thermodynamically
insignificant. We can calculate the free energy
$F= -T \ln{Z_{torus}}$,  using $N! \simeq N\ln{N} - N$ for large $N$,
\eq
     F \simeq -NT(-\ln{N} + \ln{(A- 4 \pi N)} +
\ln{\frac{2e\pi^{2}T}{h^{2}}})
\eqe
The pressure $P=-\frac{\del F}{\del A}$ is
\eq
     P =  \frac{NT}{A-4\pi N} \label{eq:pressure}
\eqe
and the entropy $S=-\frac{\del F}{\del T}$ is proportional to $N$:
\eq
S = N(-\ln{n} + \ln{(1-4\pi n)} + \ln{\frac{2e^{2}\pi^{2}T}{h^{2}}})
\label{eq:entropy}
\eqe
So, in the thermodynamic limit $N \rightarrow \infty$ at fixed $n$,
we obtain the same physical results
on the torus as on the sphere, a remarkable result given the disparity
of the mathematical methods used to arrive at the result for the two
manifolds. This result supports the general principle that in the
thermodynamic limit the
behaviour of the gas is independent of the global topology of the
physical manifold it occupies, even though topological methods
have been used to calculate the partition function. Since $A
\rightarrow \infty$ as $N \rightarrow \infty$ and since the torus is
intrinsically flat, we can interpret the results (\ref{eq:pressure})
and (\ref{eq:entropy}) as describing the thermodynamics of vortices in
the plane.
\par
In the low density limit we find
\eq
PA = NT(1+4\pi n +O(n^{2}))
\eqe
as would be obtained for a gas of hard discs of area $2\pi$, and in
the high density limit $n= \frac{1-\e}{4\pi}$, with $\e$ small
\eq
P=T \e^{-1} + O(1) \label{eq:hipress}
\eqe
This describes what would happen if, keeping the number of vortices $N$
fixed, we were to decrease the physical area available to the vortices
towards the limit $A = 4\pi N$; the work needed to compress the
surface $-\int P dA$ would diverge as $A \rightarrow 4\pi N$.
\par
We also note that the pressure and free energy are differentiable,
provided $A > 4\pi N$,  and
hence there is no phase transition in a vortex gas at critical
coupling. This is perhaps not unexpected due to the smooth
interactions of the vortices and the absence of forces between them.
\par
It would be interesting to investigate the behaviour of
the vortices some small distance away from the critical coupling
either towards a Type I superconductor (so that the vortices would be weakly
attracting) or a Type II superconductor (with vortices repelling).
In these cases, it is possible that the vortex gas could undergo a
phase transition between a gaseous and a condensed (type I) or
crystalline (type II) phase.

\newpage

\begin{figure}
\setlength{\unitlength}{0.0125in}%
\begin{picture}(430,298)(90,370)
\thicklines
\put(160,520){\oval( 40, 40)[br]}
\put(160,520){\oval( 40, 40)[tr]}
\put(160,400){\oval( 40, 40)[tr]}
\put(340,640){\oval( 40, 40)[bl]}
\put(340,520){\oval( 40, 40)[tl]}
\put(340,520){\oval( 40, 40)[bl]}
\put(340,400){\oval( 40, 40)[tl]}
\put(160,640){\oval( 40, 40)[br]}
\put(160,520){\circle{10}}
\put(337,640){\circle{10}}
\put(340,400){\circle{10}}
\put(340,520){\circle{10}}
\put(160,640){\circle{10}}
\put(160,400){\circle{10}}
\put(160,400){\framebox(360,240){}}
\put(340,620){\line( 0,-1){ 80}}
\put(180,400){\line( 1, 0){140}}
\put(160,500){\line( 0,-1){ 80}}
\put(160,620){\line( 0,-1){ 80}}
\put(180,640){\line( 1, 0){140}}
\put(340,500){\line( 0,-1){ 80}}
\multiput(165,580)(-0.25000,0.50000){21}{\makebox(0.4444,0.6667){\sevrm .}}
\multiput(160,590)(-0.25000,-0.50000){21}{\makebox(0.4444,0.6667){\sevrm .}}
\multiput(255,635)(0.41667,0.41667){13}{\makebox(0.4444,0.6667){\sevrm .}}
\multiput(260,640)(-0.41667,0.41667){13}{\makebox(0.4444,0.6667){\sevrm .}}
\multiput(335,470)(0.41667,-0.41667){13}{\makebox(0.4444,0.6667){\sevrm .}}
\multiput(340,465)(0.41667,0.41667){13}{\makebox(0.4444,0.6667){\sevrm .}}
\multiput(250,395)(-0.41667,0.41667){13}{\makebox(0.4444,0.6667){\sevrm .}}
\multiput(245,400)(0.41667,0.41667){13}{\makebox(0.4444,0.6667){\sevrm .}}
\put(255,510){\line( 1, 0){  5}}
\put(400,520){\vector(-3,-4){ 46.800}}
\put(400,520){\vector(-1, 1){ 52.500}}
\put(110,525){\vector( 1, 1){ 42.500}}
\put(110,525){\vector( 2,-3){ 40}}
\put(125,650){\line( 1, 0){ 15}}
\put(130,650){\line(-1,-3){  5}}
\put(495,375){\makebox(0,0)[lb]{\raisebox{0pt}[0pt][0pt]{\twtyrm x=1}}}
\put(240,510){\makebox(0,0)[lb]{\raisebox{0pt}[0pt][0pt]{\twtyrm U}}}
\put(255,500){\makebox(0,0)[lb]{\raisebox{0pt}[0pt][0pt]{\ninit 2}}}
\put(255,510){\makebox(0,0)[lb]{\raisebox{0pt}[0pt][0pt]{\ninit 1}}}
\put(130,375){\makebox(0,0)[lb]{\raisebox{0pt}[0pt][0pt]{\twtyrm 2N-3}}}
\put(90,400){\makebox(0,0)[lb]{\raisebox{0pt}[0pt][0pt]{\twtyrm x=0}}}
\put(330,650){\makebox(0,0)[lb]{\raisebox{0pt}[0pt][0pt]{\twtyrm 1}}}
\put(140,510){\makebox(0,0)[lb]{\raisebox{0pt}[0pt][0pt]{\twtyrm 1}}}
\put(350,510){\makebox(0,0)[lb]{\raisebox{0pt}[0pt][0pt]{\twtyrm 1}}}
\put(155,650){\makebox(0,0)[lb]{\raisebox{0pt}[0pt][0pt]{\twtyrm 2N-3}}}
\put(90,635){\makebox(0,0)[lb]{\raisebox{0pt}[0pt][0pt]{\twtyrm x=}}}
\put(335,375){\makebox(0,0)[lb]{\raisebox{0pt}[0pt][0pt]{\twtyrm 1}}}
\put(245,655){\makebox(0,0)[lb]{\raisebox{0pt}[0pt][0pt]{\twtyrm l}}}
\put(245,380){\makebox(0,0)[lb]{\raisebox{0pt}[0pt][0pt]{\twtyrm l}}}
\put(180,420){\makebox(0,0)[lb]{\raisebox{0pt}[0pt][0pt]{\twtyrm l}}}
\put(190,415){\makebox(0,0)[lb]{\raisebox{0pt}[0pt][0pt]{\ninit e}}}
\put(255,650){\makebox(0,0)[lb]{\raisebox{0pt}[0pt][0pt]{\ninit 2}}}
\put(255,375){\makebox(0,0)[lb]{\raisebox{0pt}[0pt][0pt]{\ninit 4}}}
\put( 90,515){\makebox(0,0)[lb]{\raisebox{0pt}[0pt][0pt]{\twtyrm l}}}
\put(100,510){\makebox(0,0)[lb]{\raisebox{0pt}[0pt][0pt]{\ninit 1}}}
\put(405,515){\makebox(0,0)[lb]{\raisebox{0pt}[0pt][0pt]{\twtyrm l}}}
\put(415,510){\makebox(0,0)[lb]{\raisebox{0pt}[0pt][0pt]{\ninit 3}}}
\end{picture}
\caption{Integration contour $\del U_{\half}$ and residues of $b$}
\end{figure}


\newpage

\begin{thebibliography}{99}
\bibitem{kolb} E. Kolb and M. Turner, The Early Universe (Chicago
University Press 1990)
\bibitem{skyrmions} T.H.R. Skyrme, Nucl. Phys. 31(1962) 556 ; E.M.Nyman
and D.O.Riska, Rep. Prog. Phys. 9(1990) 1137
\bibitem{geoapprox} N.S. Manton, Phys. Lett. B110(1982) 54
\bibitem{bogoeqns} E.B. Bogomolny, Sov. J. Nucl. Phys. 24(1976) 449
\bibitem{smv1} N.S. Manton, Nucl. Phys. B400(1993) 624
\bibitem{bradlow} S. Bradlow, Commun. Math. Phys. 135(1990) 1
\bibitem{gunning} R.Gunning, Riemann Surfaces and Generalised Theta
Functions (Springer Verlag 1976)
\bibitem{ww} E.T.Whittaker and G.N.Watson, A Course of Modern Analysis
(Cambridge University Press 1980)
\bibitem{samols} T.M. Samols, Commun. Math. Phys. 145(1992) 149 ; PhD
Thesis, Cambridge University 1990 (unpublished)
\bibitem{griffiths} P. Griffiths and J. Harris, Principles of
Algebraic Geometry (Wiley 1978)
\end{thebibliography}
\end{document}